# The Role of Singular Solutions in the Study of Physical Systems

Vyacheslav A. Buts

*National Science Centre "Kharkov Institute of Physics and Technology", Kharkiv, Ukraine;
Institute of Radio Astronomy of NAS of Ukraine, Kharkiv, Ukraine;*

*Abstract:* The necessity and benefit of singular solutions in the study of physical systems is shown. By singular solutions we mean solutions that are not contained in the general solution of the system of equations that describes the dynamic system under study. In addition, at the points of singular solutions the conditions of the uniqueness theorem are violated. It is shown that the presence of singular solutions, first of all, leads to the emergence of chaotic regimes. The dynamics of the system under study in the presence of singular solutions can differ radically from the dynamics in which singular solutions are not considered. It is shown that in many cases, in the presence of singular solutions, the system under study may turn out to be anomalously sensitive to small perturbations. Typically, singular solutions are not considered at analysing the dynamics of physical systems. It is shown that taking such solutions into account even in well-studied physical problems (for example, in the Kepler problem) can lead to unusual chaotic regimes. Examples of physical systems in which considering singular solutions turned out to be useful are given. It is shown that the use of an invariant measure provides a simple criterion for the appearance of singular solutions. It is shown that any regular function can be represented as a set of random functions.

## Introduction

In this work, by the term singular solutions we will understand solutions at the points of which the uniqueness conditions are violated (the Lipschitz condition is not satisfied). Such solutions are widely presented in the mathematical literature. However, when analysing mathematical models that describe physical processes, conditions are explicitly or implicitly imposed that exclude taking such decisions into account, i.e. conditions for the uniqueness of solutions are imposed. As an example, we can cite the definition of the main property of phase space, which considers the regular and stochastic dynamics of physical systems in the book [1] "At any given moment in time, trajectories in phase space do not intersect..." This proposal immediately excludes the inclusion of singular solution. In addition, this proposal immediately imposes the need to implement a regime with dynamic chaos to have a system whose number of degrees of freedom is greater than or equal to 1.5. Indeed, if we have a system with one degree of freedom, then the phase space of such a system is a plane. If phase trajectories cannot intersect, then the conditions for mixing these trajectories cannot be realized on the plane.

This attitude towards singular solutions is apparently because they do not correspond to our idea of the need to obtain certain solutions in physical problems. We also note that a detailed analysis of singular solutions was given by V.A. Steklov [2]. He also gave an algorithm for finding such solutions. However, Steklov himself, analysing the possibility of the appearance of such solutions in real physical problems, concluded that they do not describe any real physical processes. In particular, he writes: "The singular solutions of different types are not always possible for any given system of differential equations. As practice shows, singular solutions are obtained only in exceptional cases. For example, all currently resolved questions of general mechanics and astronomy are obtained by integrating equations ***that do not allow any singular solutions***. It is likely that all differential equations of mechanics do not have any singular solutions, although this proposal, which has important philosophical significance, as far as I know, remains generally unproven at the present time."

By now it has become clear that the solutions introduced by Steklov are envelopes of other solutions. That is why the uniqueness conditions are violated at the points of these solutions. In this paper we will be interested in the role namely of such solutions. Note that the number of works in which such solutions play an important role to one degree or another is very small. References to some of them are contained in the list of references.

The further text of the work contains four sections, appendix, conclusion and reference. The *second section* examines the famous Kepler problem. It is shown that this problem has singular solutions. Moreover, these solutions were obtained in two ways: the first using an algorithm that was described by Steklov [2], the second because of finding an envelope at the points of which the conditions of the uniqueness theorem are violated. It is shown that considering singular solutions makes it possible to assess the degree of reliability of the results obtained within the framework of the model under consideration. The Steklov algorithm and the envelope method are difficult to implement in many cases. The question arises about the existence of some additional features (criteria) that allow us to indicate the presence of special solutions. The *third section* of the work is devoted to this issue. It is shown that the use of an invariant measure provides a simple criterion for the possibility of the emergence of singular solutions in the problem under consideration.

In the *fourth section*, the well-known physical problem of three-wave interaction of waves is considered. This problem was chosen because in obtaining a system of equations that describe such an interaction, a replacement was used, which leads to the emergence of singular solutions. Note that this replacement is very widely used in physical models. Its popularity is because it allows one to distinguish the amplitude of a process (usually a slow process) and phase (a fast process).

When studying any physical problems, a key role is played by conserved quantities - integrals. In modes with dynamic chaos, these integrals are determined by certain expressions for the dependent variables, which are random functions. This fact suggests the idea that any regular function can be represented as a combination of random functions. This is the subject of the *fifth section*.

The algorithm for finding singular solutions does not seem to be described in the physical literature. Therefore, *Appendix* provides the key elements of this algorithm. It seems that this algorithm can only be successfully used with the help of PC.

## 2. Kepler problem

First of all, we want to understand how singular solutions arise, what the emergence of singular solutions leads to, what new features of the dynamics of the systems under study appear? The easiest way to do this is with a simple example. As such an example, we can choose the Kepler problem. Indeed, this problem is quite simple, has been studied in detail and, as we will see below, its solutions contain singular solutions.

As the main expression that can be used to study the dynamics of the motion of bodies in the Coulomb (gravitational) potential, we will choose the expression for the energy of the body whose dynamics are being studied:

$$E = \frac{m}{2}\dot{r}^2 + U_{eff}(r) = const \ . \quad (1)$$

Where - $U_{eff} = U + \frac{m}{2}r^2\dot{\varphi}^2 = U + \frac{M^2}{2mr^2}$,

$M = (mr\dot{\varphi})r = const$ - angle momentum.

We use this integral to obtain a system of differential equations. To do this, we introduce the Coulomb (for the movement of charged particles) or gravitational (Kepler problem) potential: $U = -Q/r$. In addition, for further purposes it is convenient to enter the following parameters: $M/m = a$; $Q/m = b$.

$$\dot{r} = v$$
$$\dot{v} = -\frac{1}{m}\frac{\partial U_{eff}}{\partial r} + \beta\cos(\omega t) = (\frac{a^2}{r^3} - \frac{b}{r^2}) + \beta\cos(\omega t) \ . \quad (2)$$
$$\dot{\varphi} = M/mr^2 .$$

An additional term $\beta\cos(\omega t)$ is introduced to the right side of the second equation, which describes the influence of external disturbances. Sometimes it is more convenient to consider an equivalent system of equations instead of this one (without additional perturbation $\beta = 0$):

$$\dot{r} = \pm\sqrt{\frac{2}{m}[E - U(r)] - \frac{M^2}{2mr^2}}$$
$$\dot{\varphi} = M/mr^2 \quad (3)$$

For the Coulomb (gravitational) potential, the ODE system (3) has a simple analytical expression for the general solution. This solution expresses the relationship between the radius and the angular variable. Let us write this solution in the form presented by Landau [3]:

$$r = p/[1 + e\cos(\varphi + C)] \ . \quad (4)$$

Here $p = M^2/mQ$ -parameter, $e = \sqrt{1+(2EM^2)/mQ^2}$ - eccentricity.

Let's use the procedure for finding special solutions, which is described by Steklov [2] (see Appendix). For this case, we find the following values for the "constant" C:

$$C = -\varphi + n\pi. \quad (5)$$

Substituting this expression into the general solution (4), we find the following expressions for singular solutions:

$$r_s = p/[1 \pm e]. \quad (6)$$

These solutions are not contained in the general solution (4). However, they satisfy equations (1) at $\dot{r}=0$. In addition, it is easy to see that the Lipschitz condition is violated at the points of these solutions. Therefore, these solutions are singular solutions. It is also easy to see that the points of singular solution for captured particles ($r_{min} \leq r \leq r_{max}$) are cusp points. Thus, the dynamics of particles at these points should acquire some features. In many (if not all) cases, singular solutions are envelopes of families of general solutions. Let's use this fact and find singular solutions as envelopes.

$$F(r,\dot{r},t) = E + \frac{Q}{r} - \frac{M^2}{2mr^2} - \frac{m}{2}\dot{r}^2 = 0. \quad (7)$$

Let us also consider its derivative with respect to $\dot{r}$ and equate it to zero

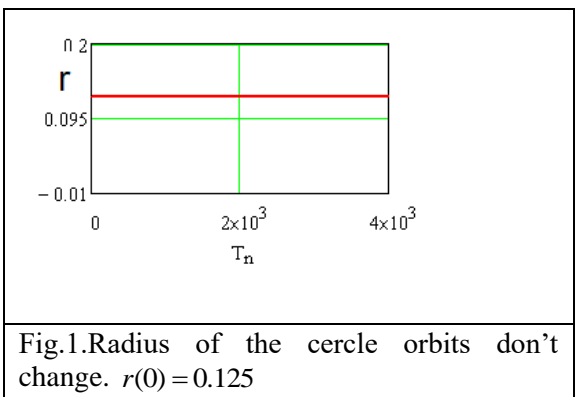

Fig.1. Radius of the cercle orbits don't change. $r(0) = 0.125$

$$\frac{d}{d\dot{r}} F(r,\dot{r},t) = -m\dot{r} = 0 \Rightarrow \dot{r}=0. \quad (8)$$

Let's substitute (8) into (7):

$$\left[E - U(r) - \frac{m}{2}r^2\dot{\varphi}^2\right] = E + \frac{Q}{r} - \frac{M^2}{2mr^2} = 0. \quad (9)$$

Let us consider that $E<0$. Then we will rewrite the equation for the envelope in a form convenient for us:

$$r = \frac{Q}{2|E|}[1 \pm e]. \quad (10)$$

Easy to see that (10) is the same as (6).

Let's show that on the points of this solution the condition of the uniqueness theorem is violated.

In this case from (1) one can see:

$$\dot{r} = \sqrt{\frac{2}{m}} \sqrt{\left[|E| - \frac{Q}{r} + \frac{M^2}{2mr^2}\right]} = f(r). \quad (11)$$

$$\frac{\partial f}{\partial r} = \frac{1}{2}\sqrt{\frac{2}{m}} \frac{(Q/r^2 - M^2/mr^3)}{\sqrt{\left[|E| - \frac{Q}{r} + \frac{M^2}{2mr^2}\right]}}.$$

The denominator, under the root, contains equation (9), which describe envelope. So, at the points of our interest the conditions Picar theorem are violated.

**2.1 Sensitivity to change initial position.**

We will analyse the dynamics using numerical methods. The degree of accuracy of numerical calculations will be characterized by the values of the convergence of lines in the distance (TOL) and constrained tolerance (CTOL).

The variety of numerical results is very large. Therefore, the most characteristic and important ones will be given below. First, we note that the dynamics in strictly circular orbits do not feel the influence of singular decisions. But in these orbits the eccentricity is zero. However, when the initial conditions deviate from a circular orbit, the presence of singular solutions can qualitatively change the dynamics. This fact is shown in Figures 1-6. All calculations were carried out at TOL=CTOL=0.001. In addition, all calculations were carried out with the same values of the following parameters: $a = 0.5; b = 2$. The difference was only in the initial value of the radius: $r(0) = 0.125$ for Fig.1, Fig.3, Fig.5, and $r(0) = 0.2$ for Fig.2, Fig.4, Fig.6

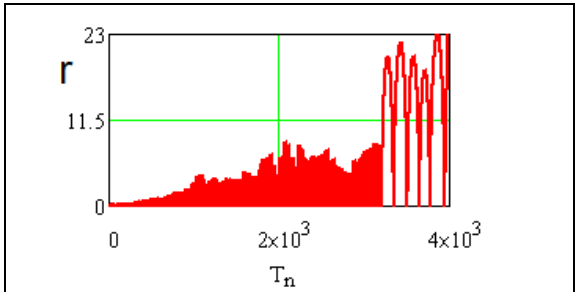

*Fig.2. Changing the initial conditions $r(0) = 0.2$ qualitatively changes the particle dynamics*

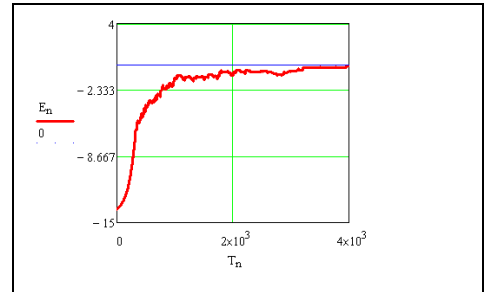

*Fig.6. Energy change in the presence of special solutions. $r(0) = 0.2$*

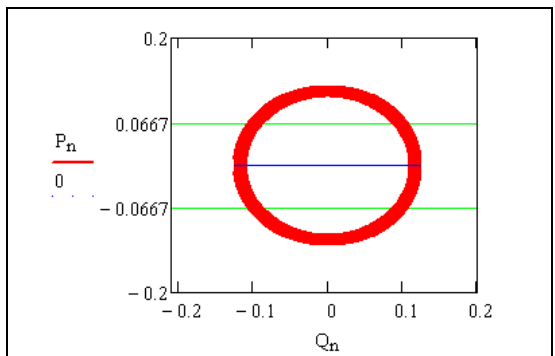

*Fig 3. Here $r(0) = 0.125$, Circular orbit $P = r \cdot \cos\varphi; \; Q = r \sin\varphi$*

It should be noted that in the system of equations (2) the third equation is isolated from the first two. Thus, the dynamics in the Kepler problem is described by a system that has only one degree of freedom. In this respect, the situation is like that studied in the works [4-6]. However, it is known that the Poincare-Bendixson theorem prohibits the occurrence of chaotic regimes in systems with one degree of freedom. The question arises. What is the problem? The answer is as follows. In the Kepler problem, such regimes arise for two reasons. The first reason is the presence of singular solutions, and the second is that the numerical algorithms used have their own noise (numerical noise). Indeed, if we solve the system of equations (2) using more accurate algorithms, the chaotic modes will disappear. To demonstrate this statement the figure 7 show the correlation function of particle motion in the Kepler problem with a high degree of calculation accuracy ( TOL=$10^{-6}$). It is evident that the correlation function oscillates without changing its amplitude.

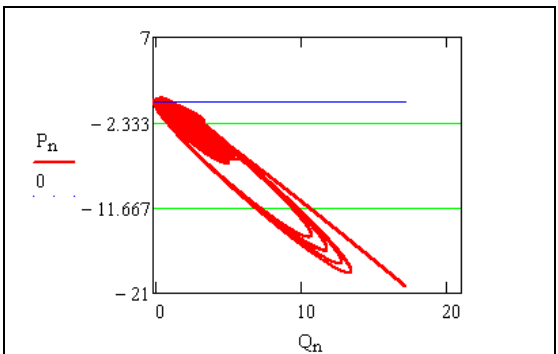

*Fig.4. The same as in Fig.3. But $r(0) = 0.2$*

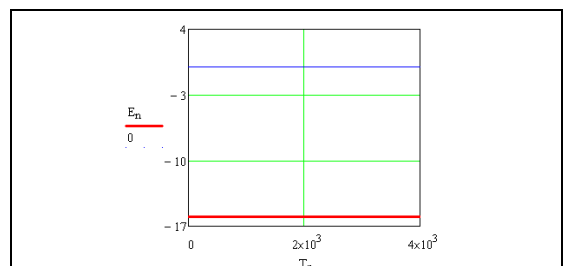

*Fig.5. Energy at cercle orbits $r(0) = 0.125$ don't change.*

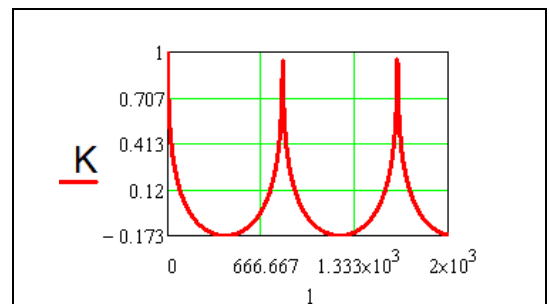

***Fig.7.*** *Autocorrelation of the function $x_0$ at high level calculation (TOL=$10^{-6}$). $\beta = 0$*

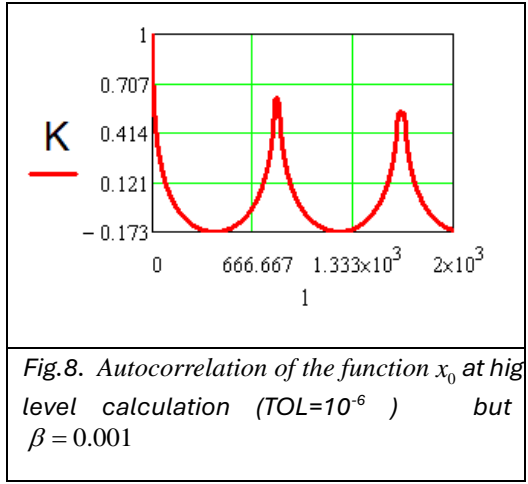

*Fig.8. Autocorrelation of the function $x_0$ at high level calculation (TOL=$10^{-6}$) but $\beta = 0.001$*

However, the fact that the Kepler problem has singular solutions means that about these solutions the density of integral curves is very high. Therefore, it can be expected that this system will be very sensitive to small external disturbances and to the appearance of chaotic regimes. To prove the existence of such a feature, we had introduced a small regular, periodic disturbance $\beta \cos \omega t$ into the right-hand side of the second equation of system (2). The presence of this disturbance makes the system (2) non-autonomous and having 1.5 degrees of freedom. In addition, the dynamics of particles in the Kepler problem in this case becomes irregular even with a high degree of accuracy of the calculations. The figure 8 shows the correlation function of the Kepler problem in the presence of a small disturbance ($\beta = 0.001$). It is evident that the amplitude of the correlation function decreases. It should be noted that more accurate numerical calculations of the systems (TOL $\leq 10^{-6}$), which are described in the works [4-7] also do not show the presence of chaotic regimes.

Let's notice that the figures 7 and 8 were constructed with the following parameter values $\omega = 0.05$, $r(0) = 5$ $a = \sqrt{0.05}$, $b = 1$, $\beta = 0$ for Fig.7 and $\beta = 0.001$ for Fig.8.

## 3. Measure - as a criterion for the emergence of chaotic dynamics

The examples discussed above are quite simple. We knew the analytical form of the integral curves of these systems. This fact allowed us to identify solutions in which the uniqueness theorem does not hold. In more complex cases, such a possibility arises quite rarely, so one would like to find simpler and more general criteria that will allow us to determine areas of phase space in which elements of unpredictability may appear. One possibility is to use a measure. Indeed, let us introduce the measure of the "segment" $\Delta \vec{x}$: $\Delta \mu = p(\vec{x}_i) \cdot \Delta \vec{x}$. Here $p(\vec{x}_i)$ - probability density of finding a representative point in this segment. Let, because of the time dynamics of the system under consideration, the point $\vec{x}_i$ transforms to point $\vec{z}$.

Thus, we have $\vec{z} = f(\vec{x}_i)$ - the image of the point $\vec{x}_i$; $\vec{x}_i$ - preimage of a point $\vec{z}$. There can be many prototypes. Let us now consider a certain "segment" $\Delta \vec{z}$: ($[\vec{z} - \Delta \vec{z}/2; \vec{z} + \Delta \vec{z}/2]$). The measure of this segment will now be determined by the formula:

$$\Delta \mu_z = g(\vec{z}) \cdot \Delta \vec{z} = \sum_i p(\vec{x}_i) \cdot \Delta \vec{x}_i . \qquad (12)$$

Here $g(\vec{z})$ is the probability density of finding the representing point in the phase volume $\Delta \vec{z}$. From this formula we find the expression for density

$$g(\vec{z}) = \sum_i p(\vec{x}_i) \frac{\Delta \vec{x}_i}{\Delta \vec{z}} = \sum_i \frac{p(\vec{x}_i)}{|J_i|} . \qquad (13)$$

Here $J_i$ is the Jacobian of the transformation of new variables through old variables.

Formula (13) is practically the Perron-Frobenius formula. From this formula in those areas where the Jacobian of the transformation will have some features (for example, it will turn to infinity or zero), we can expect that the relationships between the original probability densities and the transformed ones will become uncertain. These areas can be sources of chaos.

Let's look at the most used and most important replacement as an example. Let there be a dependent variable $\vec{x} = (A', A'')$. Here $A = A' + iA''$. Let's make a replacement: $A = A' + iA'' = a \cdot \exp(i\varphi)$. Here $a$ is the modulus of the complex amplitude: $A$: $a = \pm\sqrt{A'^2 + A''^2}$, $\varphi = arctg(A''/A')$ - phase of this variable. It's easy to check what in this case $J = 1/a$. Thus, if $a \to 0$, then we find ourselves in the area of unpredictability. This result corresponds to the result from the previous section that under such transformations the points of the phase space in which $a \to 0$ correspond to the points of singular solutions.

Note also that the considered transformation describes the transition from quantum consideration to classical consideration. In this case, the condition

$a \to 0$ corresponds to the tendency of the particle velocity to zero. This fact corresponds to the well-known result that transitions from quantum consideration to classical consideration for particles with zero velocities are not correct

## 4. Dynamics of three-wave interaction

The second, well-studied physical process, the dynamics of which can be significantly affected by considering singular solutions, is the process of nonlinear three-wave interaction. When formulating a mathematical model of such interaction, the replacement of dependent variables is often used in the form:

$$A_k = a_k \exp(i \cdot \varphi_k). \qquad (14)$$

Here $a_k$ is the modulus of the dependent complex variable; $\varphi_k$ is the real phase of this variable. This replacement is very widespread. Examples of such transformations can be the transformations that are carried out during the transition from the equations of quantum mechanics to the Hamilton-Jacobi equations of classical mechanics. In addition, such transformations are widely used in radiophysics, electronics, and plasma physics. In all these cases, systems of equations that describe the process under study in new variables contain regions in which the uniqueness theorem is violated. Below, using the example of nonlinear interaction of three waves in nonlinear media (in particular, in plasma), we will show what features can arise in this case.

To be specific, we will follow the results presented in [8]. In this monograph, a system of equations is obtained that describes the dynamics of the complex amplitudes of three interacting waves in a plasma. This system of equations can be represented as follows:

$$\dot{x}_0 = -x_1 x_2 \cos x_3 \ ; \quad \dot{x}_1 = x_0 x_2 \cos x_3 \ ;$$

$$\dot{x}_2 = x_1 x_0 \cos x_3 \ ;$$

$$\dot{x}_3 = \Delta\omega - \left[ \frac{x_0 x_1}{x_2} + \frac{x_0 x_2}{x_1} - \frac{x_2 x_1}{x_0} \right] \sin x_3 + \delta \cos \omega t \ . \qquad (15)$$

Here $x_k = a_k$; $k = \{0;1;2;\}$ - modules of complex amplitudes of interacting waves;

$x_3 = \varphi_0 - \varphi_1 - \varphi_2 + \Delta\omega \cdot t$; $\varphi_k$; $k = \{0;1;2\}$ - real phases of interacting waves. If $\Delta\omega = 0$; $\delta = 0$, then the system of equations (15) has the following integrals:

$$x_0^2 + x_1^2 = M_1 = const \ ; \ x_0^2 + x_2^2 = M_2 = const \ . \qquad (16)$$

The system of equations (15) for $\delta = 0$ has been studied in detail in many papers and monographs. Below we will be interested in the influence of precisely this small additional disturbance on the dynamics of wave interaction. From system (15) it is immediately clear that at points in the phase space where the magnitudes of the amplitudes of the interacting waves tend to zero, the uniqueness theorem will be violated. In addition, from the same system the coordinates of these points are solutions to system (15). Singular solutions. In the real dynamics of wave interaction, phase trajectories only approach these points. The situation is largely like that described in the previous section regarding the Kepler problem. When phase trajectories pass near these points, insignificant, unaccounted forces can lead to a radical change in dynamics. These features must be kept in mind. To illustrate this influence of small forces, we consider the dynamics of the system of equations (15) in the presence of a small periodic disturbance. We will compare the dynamics of system (15) in the absence of a small disturbance and in its presence. We will choose the initial conditions for all cases in the unstable zone (in the region of parameters where the decay process is most pronounced):

$x_0 = 0.999$; $x_1 = 0.01$; $x_2 = 0.044$; $x_3 = -0.01$; $\Delta\omega = 0$; ( TOL=CTOL=$10^{-5}$).

First, we will consider the case when there is no disturbance ($\delta = 0$). The results of a numerical study of system (15) in this case are presented in Figures 9-11. These figures show the characteristic dynamics of changes in the moduli of interacting amplitudes, as well as their statistical processing (spectra and autocorrelation function). The well-known regular time dynamics of interacting waves is visible: the amplitude magnitudes change periodically. The spectra are narrow, the correlation function oscillates, but its maximum amplitude does not change. For the chosen initial values, the amplitudes of the interacting waves approach quite closely to the region where the uniqueness theorem is violated. It can be expected that if the accuracy of numerical calculations is insufficient, the amplitudes will fall into these regions and, as a result, the dynamics may become irregular. Indeed, such a process takes place. However, we will focus our attention on the case when the calculation accuracy is high, but the system can be affected by a small external force that is not considered in the model. ***This situation is of the greatest interest, since a high degree of numerical calculations can lead to unfounded confidence that we obtain the correct dynamics in our studies. We will see that very small external forces can seriously disrupt this dynamic.***

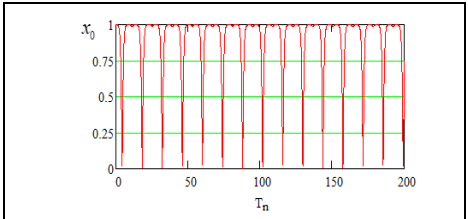

*Fig.9. Amplitude dependence decaying wave over time*

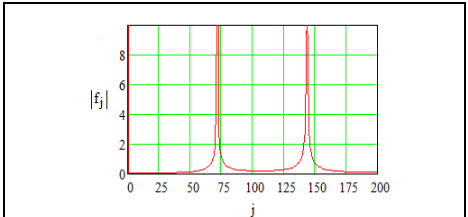

*Fig10. The spectrum of the variable $x_0$*

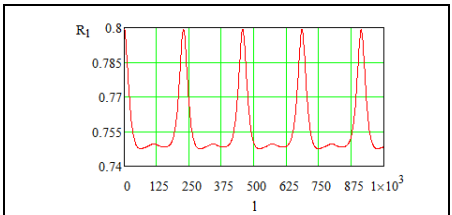

*Fig.11. Autocorrelation of the function $x_0$*

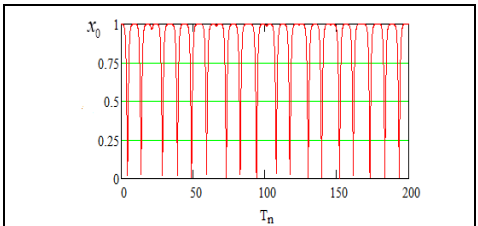

*Fig.12. The dependence of the decaying wave amplitude over time*

Let us illustrate the above by calculating the system of equation (15) considering a small external disturbance. We will assume that the amplitude of this disturbance and its frequency are equal $\delta = 0.001; \omega = 1$. The calculation results for this system are shown in Figures 12-14. From these figures it is clear that the dynamics have already become less regular. The spectra have expanded significantly, and the correlation function decreases quite quickly. All this suggests that under these conditions, considering even insignificant external additional forces can lead to a qualitative change in the dynamics of wave interaction.

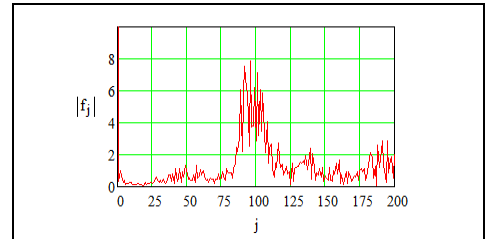

*Fig.13. Spectrum variable $x_0$ in the presence of disturbance*

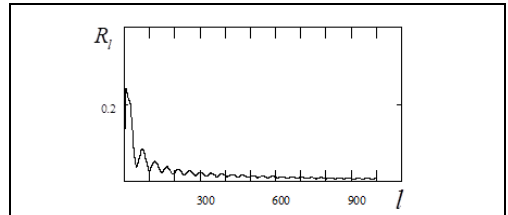

*Fig.14. Autocorrelation function of a variable $x_0$ in the presence of disturbance*

## 5. Decomposition of regular functions into a combination of random functions.

In this section, using a simple example, it is shown that almost any regular function can be represented as some combination of functions whose dynamics are chaotic. To do this, consider the following system of ordinary differential equations:

$$\dot{x} = y, \quad \dot{y} = -x \quad . \qquad (17)$$

This system represents the equations of a linear oscillator. We will be interested in the case when these variables represent complex functions. We want to find such replacements of variables that will enable us to determine the dynamics of the real and imaginary components of these variables. In addition, we will be interested in such substitutions that lead to the emergence of chaotic dynamics. If we use the trivial replacement: $x = x_R + i \cdot x_I$, $y = y_R + i \cdot y_I$, Then, how easy it is to see, that the real and imaginary components are separated. Really:

$$\dot{x}_R = y_R, \quad \dot{y}_R = -x_R. \qquad (18)$$

$$\dot{x}_I = y_I, \quad \dot{y}_I = -x_I . \quad (19)$$

In this case, the dynamics of the real and imaginary parts represent the dynamics of a linear oscillator. No new dynamics arise. New dynamics arise when the dependent variables are changed differently. Namely, when we will use replacing:

$$x = x_0 \cdot \exp(i \cdot x_2) \quad y = x_1 \cdot \exp(i \cdot x_3) . \quad (20)$$

Here $x_k(t)$ - real functions; $k = \{0,1,2,3,\}$

Substituting (20) into (17), we obtain the following system of equation for finding new dependent variables:

$$\dot{x}_0 = x_1 \cos(x_3 - x_2), \quad \dot{x}_1 = -x_0 \cos(x_3 - x_2),$$

$$\dot{x}_2 = \frac{x_1}{x_0} \sin(x_3 - x_2), \quad \dot{x}_3 = \frac{x_0}{x_1} \sin(x_3 - x_2) . \quad (21)$$

These four equations can be rewritten as three equations:

$$\dot{x}_0 = x_1 \cos \Phi, \dot{x}_1 = -x_0 \cos \Phi,$$

$$\dot{\Phi} = \left( \frac{x_0}{x_1} - \frac{x_1}{x_0} \right) \cdot \sin \Phi . \quad (22)$$

Here $\Phi = x_3 - x_2$.

In addition, from the first two equations of system (21) or (22) these systems have the following integral:

$$x_0^2 + x_1^2 = const , \quad (23)$$

Thus, the system of equations (22) has only one degree of freedom. The original system (17) also had one degree of freedom. Let's return to (21). It is easy to see that this system has a solution $x_0 = x_1 = 0$. This is a singular solution (at the points of this solution the conditions of the uniqueness theorem are violated). In accordance with (23), such a solution will be realized only under zero initial conditions. Under other initial conditions, the system will only periodically fall into singular solution points. As soon as it hits a point, for example $x_0 = 0$, the corresponding phase ($x_2$) becomes uncertain. It randomly either does not change or changes (in a jump) by an amount $\pi$. This feature of the dynamics of amplitude and phase is demonstrated in Figures 15,16. Figure 15 shows that when the zero value passes, the derivative of the dependent variable either does not change or goes to zero. In accordance with this feature of the amplitude dynamics, at the same moments of time, the phase $x_2$ either does not change or changes by an amount $\pi$ (see Fig. 16).

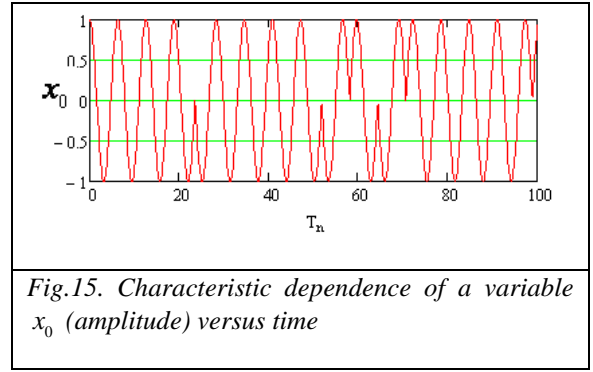

*Fig.15. Characteristic dependence of a variable $x_0$ (amplitude) versus time*

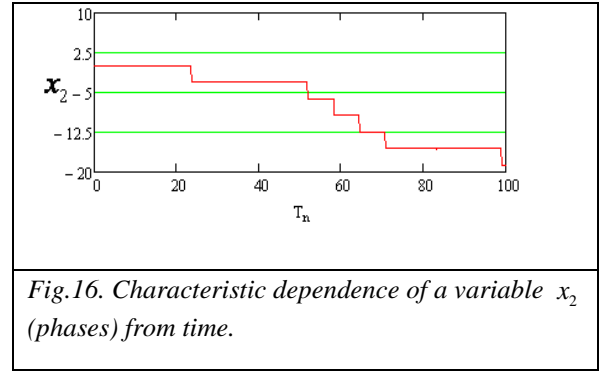

*Fig.16. Characteristic dependence of a variable $x_2$ (phases) from time.*

Looking at Figures 15 and 16 One can see that phase jumps correspond to the points where the derivative of the amplitude goes to zero as it passes the zero point. It can also be seen that the height of the steps is exactly equal $\pi$. This feature in the dynamics of the passage of the zero point of the amplitude is in full accordance with the fact that, as shown in the previous section, uncertainty is born at these points. The system randomly selects possible paths for further dynamics. In our simplest case, there are only two such paths: either continuous dynamics or dynamics with a phase jump by $\pi$. It should be said that the choice in all cases is random. It depends on the accuracy of the research performed. The slightest change in the parameters of numerical calculations changes the location of the points at which the dynamics change. The fact that these jumps are random is reflected in the behaviour of the autocorrelation function for both amplitude and phase. A characteristic form of the autocorrelation function for amplitude is shown in Fig. 17.

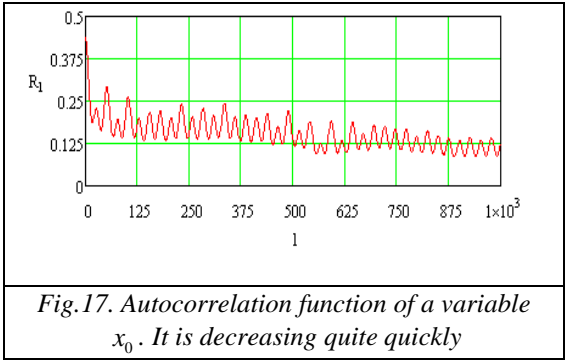

*Fig.17. Autocorrelation function of a variable $x_0$. It is decreasing quite quickly*

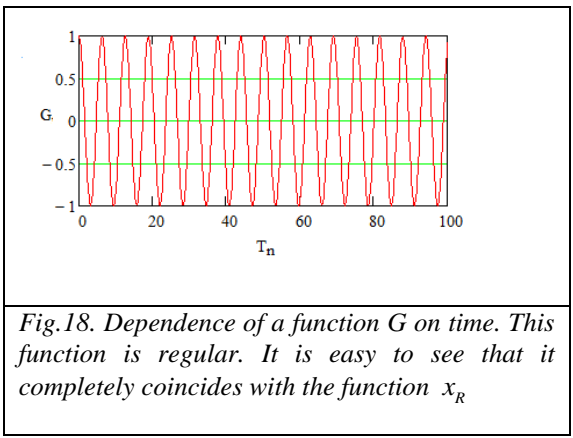

*Fig.18. Dependence of a function G on time. This function is regular. It is easy to see that it completely coincides with the function $x_R$*

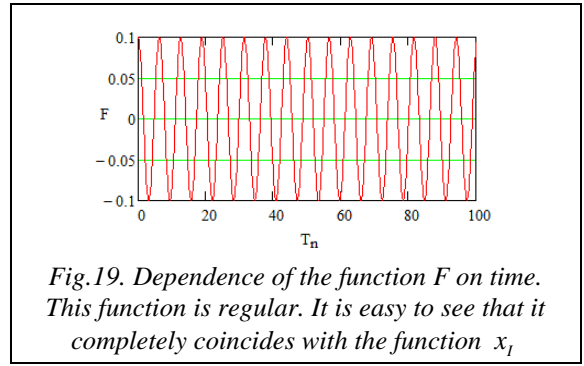

*Fig.19. Dependence of the function F on time. This function is regular. It is easy to see that it completely coincides with the function $x_I$*

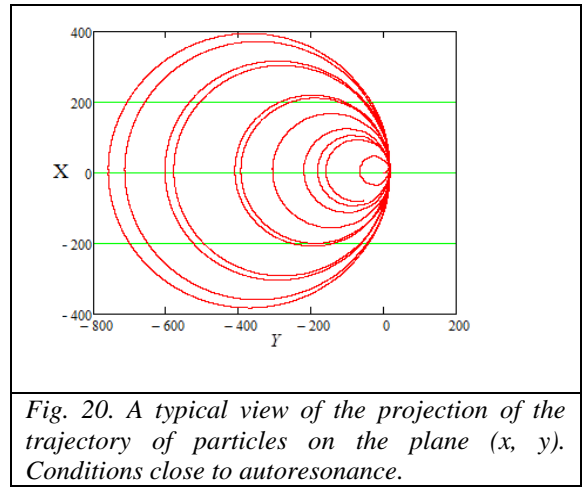

*Fig. 20. A typical view of the projection of the trajectory of particles on the plane (x, y). Conditions close to autoresonance.*

The most important and interesting fact for us is that, despite the random dynamics of the new dependent variables, some combination of these variables behaves quite regularly. Indeed, if all the replacements made and the numerical analysis carried out are quite adequate to each other, then from the new dependent variables we can construct a function that completely coincides with the original dependent variables: $F = x_0 \cdot \sin(x_2) \equiv x_I$, $G = x_0 \cdot \cos(x_2) \equiv x_R$.

Such a combination should have regular dynamics. In this case, even though the variables $x_0$ $x_2$ behave randomly, the functions $G$ и $F$ must be regular. Moreover, their dynamics must coincide with the dynamics of functions $x_R$ and $x_I$. Namely such dynamics of these functions is observed. This fact is illustrated in Figures 18 and 19. It is easy to see that this dynamic completely coincides with the dynamics of the functions $x_R$ and $x_I$. Note that the above calculations were carried out under the following initial conditions: $x_0 = 0.99999$, $x_1 = 1.999 \cdot 10^{-3}$, $x_2 = 0.1$, $x_3 = 0$

Thus, we see that even the simplest regular function $\sin(\omega t)$ or $\cos(\omega t)$ can be represented as a product of two functions, the dynamics of which is chaotic. This fact can be called hidden dynamics. It is also easy to see that much more complex regular functions can also be represented as some combination of functions with random dynamics. It should be noted that this result is in some sense not unexpected. Indeed, we know many physical systems whose dynamics are chaotic (dynamic chaos), but some combination of variables of this system turns out to be completely regular or even simply constant. In the latter case, we simply deal with the integral of the physical system being studied.

## 6. Conclusion

The most important conclusion that can be drawn from the results obtained above is that knowledge of singular solutions is necessary when studying the dynamics of almost any problem. Indeed, we had seen above that the presence of singular solutions in the Kepler problem can lead to the appearance of chaotic regimes. This may be an erroneous result. In addition, the presence of singular solutions leads to the appearance of regions in the phase space, the density of integral curves of which is very high.

Therefore, even small external disturbances (even periodic ones) can lead to chaotic regimes. Systems with only one degree of freedom are easy to understand (thanks to the Poincaré-Bendixson theorem). More complex problems are more difficult to understand.

All this indicates that when analysing the dynamics of almost any system, it is necessary to examine the system for the presence of special solutions. If such solutions exist, then it is necessary to be careful and check the results, especially numerical calculations, for the effect of the results on the degree of accuracy of the calculations, as well as for the stability of the results from the influence of small external disturbances (considering small, discarded terms).

Some simple examples were considered in detail above. They showed the important role of considering singular solutions. The number of examples could have been increased. There are not many of them. Let us point them out. In [9] it was shown that with certain changes of variables (as were used above) chaotic regimes can also arise in linear systems. In [10] it was shown that chaotic regimes with intermittency, arising in cyclotron resonances, also arise when singular solutions appear.

In addition, in the dynamics of some complex physical systems, processes are observed that can be explained by the presence of special solutions. Thus, in the work [10] dynamics of charged particles that are in an external magnetic field and in the field of external intense electromagnetic waves, processes are observed that can be qualitatively explained as result of presence of singular solutions. See figure 20. This figure shows the particle trajectories in cross section. It is evident that all trajectories have a common point (envelope). This feature of the trajectories is most characteristic of special solutions. It is also necessary to pay attention to the role of point mappings that have a singularity [11-13]. Such singular point mappings, based on the results of their calculations, can also be attributed to the modelling of singular solutions.

The importance of finding special solutions is that their presence requires a more careful analysis of the results obtained. Indeed, in the presence of such solutions, even very small disturbances not considered in the modelling can lead to dynamics that differ from the expected dynamics. For example, it may turn out that the sensational results of [14] may be caused namely by the fact that singular solutions were not considered in the modelling.

When searching for special solutions in the problem under consideration, Lipschitz criteria and invariant measure analysis (section three) can be useful. It is also useful to keep in mind that special solutions are envelopes.

# Appendix

To illustrate the difficulty of finding special solutions, let's consider one of the possible algorithms for finding them [2]. To do this, we write the system of equations in canonical form:

$$\frac{dx_k}{dt} = f_k(t, \vec{x}) \ . \tag{A1}$$

Let's assume that we have found a general solution to this system of equations. Let this solution look like:

$$x_k = \varphi_k(t, \vec{C}) . \tag{A2}$$

Already at this first step in most real cases we will encounter difficulties. Now we need to find a vector of arbitrary constants $\vec{C}$. We will assume that these constants are functions of time. Then the total time derivative of the general solution (A2) will have the form:

$$\frac{dx_k}{dt} = \frac{\partial x_k}{\partial t} + \sum_{i=1}^{n} \frac{\partial \varphi_k}{\partial C_i} \frac{dC_i}{dt} .$$

From this expression you can see that if the second term on the right side becomes zero:

$$\sum_{i=1}^{n} \frac{\partial \varphi_k}{\partial C_i} \frac{dC_i}{dt} = 0 \qquad k = \{1, 2, \ldots n\} . \tag{A3}$$

Then the general solution (A2) remains a solution to the original equation (A1) despite the dependence of the constants on time. ***This is a key point in the algorithm for finding special solutions.*** System of equations (A3) will be the main one for determining the vector of constants $\vec{C}$. First, we note that system (A3) has a nontrivial solution for derivatives $dC_i/dt$ only in the case when its determinant vanishes:

$$D \left| \frac{\varphi_1, \varphi_2, \ldots \varphi_k}{C_1, C_2, \ldots C_k} \right| = 0 \ . \tag{A4}$$

This will be the second relation for finding the vector of constants $\vec{C}$. Suppose that, using (A4), we expressed one of the constants in terms of other constants:

$$C_n = F(t, C_1, C_2, \ldots C_{n-1}) . \quad (A5)$$

Then its derivative will have the form:

$$\frac{dC_n}{dt} = \dot{C}_n = \frac{\partial F}{\partial t} + \sum_{i=1}^{n-1} \frac{\partial F}{\partial C_i} \dot{C}_i . \quad (A6)$$

Let's substitute expressions (A5) and (A6) into the first $(n-1)$ equations (A3):

$$\sum_{i=1}^{n-1} \frac{\partial \varphi_k}{\partial C_i} \frac{dC_i}{dt} + \frac{\partial \varphi_k}{\partial C_n} C'_n =$$
$$= \sum_{i=0}^{n-1} \frac{\partial \varphi_k}{\partial C_i} \frac{dC_i}{dt} + \frac{\partial \varphi_k}{\partial C_n} \left[ \frac{\partial F}{\partial t} + \sum_{i=1}^{n-1} \frac{\partial F}{\partial C_i} \frac{dC_i}{dt} \right] = 0 \quad (A7)$$
$$k = \{1, 2, \ldots \ldots n-1\}$$

System of equations (A7) is a linear system for finding $dC_i/dt$ derivatives. Solving this system, we find:

$$\frac{dC_i}{dt} = F_i(t, A_1, A_2 \ldots A_{n-1}) , \quad i = \{1, 2, \ldots \ldots n-1\} . \quad (A8)$$

Here $\vec{A}$ is a vector of new independent constants.

Formulas (A5) and (A8) allow us to potentially find all components of the vector $\vec{C}$. Substituting them into the expression for the general solution (A2), we find many special solutions. As we can see, the implementation of the considered algorithm for finding special solutions is, in the general case, a complex problem. Only in rare cases can this algorithm be implemented analytically.